\documentclass[aps,prb,twocolumn,groupedaddress]{revtex4}
\usepackage{graphicx}

\begin{document}

\newcommand{\be}{\begin{equation}}
\newcommand{\ee}{\end{equation}}
\newcommand{\bea}{\begin{eqnarray}}
\newcommand{\eea}{\end{eqnarray}}
\newcommand{\nt}{\narrowtext}
\newcommand{\wt}{\widetext}

\title{Massive Dirac fermions in single-layer graphene}

\author{D. V. Khveshchenko}

\affiliation{Department of Physics and Astronomy, University of North Carolina, Chapel Hill, NC 27599}

\begin{abstract}
Motivated by the results of recent photoemission and tunneling studies, we discuss potential many-body sources of a finite gap in the Dirac fermion spectrum of graphene. Specifically, we focus on the putative Peierls- and Cooper-like pairing instabilities which can be driven by sufficiently strong Coulomb and electron-phonon interactions, 
respectively. Our results compare favorably with the available experimental and Monte Carlo data.   

\end{abstract}

\maketitle

{\bf Introduction}

The recent advent of graphene has brought about a plethora of theoretical studies of the emergent behaviors and new phases of electronic matter in degenerate semimetalic systems. 

Amongst other spectacular experimental findings, recent 
photoemission \cite{arpes} and tunneling \cite{andrei} 
data obtained, correspondingly, in epitaxially grown and suspended graphene
demonstrated the presence of a sizable 
gap in the (othewise, nearly linear) spectrum of the Dirac-like quasiparticles which dominate 
the transport and optical properties of this novel material. Moreover, comparable  
gaps were observed in bi- and tri-layers.

Thus far, the proposed explanations of these findings invoked the effect of a commensurate static corrugation (surface reconstruction, "frozen phonon") and/or breaking of the symmetry between two sublattices (hereafter, A and B) of the graphene sheet due to a nearby substrate, which mechanisms 
can result in a hybridization of the electronic states from the vicinity of the two Dirac points (hereafter, $L$ and $R$) \cite{arpes,andrei,theory}.

Alternatively, the gap can also be generated by the Coulomb or phonon-mediated interactions, thus resulting from an incipient instability of the nodal fermion system itself.

In particular, the pertinent Coulomb interaction-driven Peierls-like instabilities can be divided onto two main groups. The first one is represented by various graphene-specific modifications of the standard scenario 
of excitonic insulating behavior developing in fermion systems with a finite Fermi surface \cite{kopaev}.

Specifically, such proposals were discussed in the cases 
of a parallel field \cite{tsvelik,lee} 
and an electrically biased bi-layer \cite{macdonald}, under either of which conditions one has matching pairs of electron and hole Dirac gases with finite (and equal - in the
case of undoped graphene) 
chemical potentials, as measured from the Dirac point.

In these BCS-type scenarios, an opening of the excitonic gap at arbitrarily weak couplings is facilitated by a finite density of states (DOS) at the Fermi surface, the gap being proportional to the effective Fermi energy of either of the two mutually attracting components. 
In particular, the authors of Ref.\cite{tsvelik} found the excitonic gap to scale (almost) linearly with the Zeeman splitting, $\Delta\sim B$.

In contrast to these scenarios, the possibility of excitonic pairing 
between the Dirac fermions with a vanishing DOS was first discussed in the context of multi-layered HOPG graphite, and later graphene,
for both, the physical long-range Coulomb interaction \cite{dvk,kiev} as well as its short-range surrogates \cite{hubbard}. By and large, the justification for using the latter, thus far, has been their computational convenience, as compared to the genuine Coulomb case.

The early analysis carried out in Refs.\cite{dvk,kiev}
predicts the existence of a finite threshold for the strength of the Coulomb coupling quantified in terms of the analog of the "fine structure constant" 
\be
g={e^2\over \epsilon v}\approx {2.16\over \epsilon}
\ee 
where $\epsilon$ is the dielectric constant of the surrounding medium (in the case of a graphene monolayer sandwiched between two different media it equals the average of the corresponding dielectric constants).

Moreover, Refs.\cite{dvk,kiev}
also discussed a concomitant phenomenon of 'magnetic catalysis', according to which a perpendicular magnetic field can open up a gap $\sim {\sqrt {B-B_0}}$, where $B_0$ is proportional to the electron density of doping relative to the Dirac point, at arbitrarily small values of $g$ \cite{dvk,kiev}. 
Unlike the mechanism proposed in Ref.\cite{tsvelik}, 
the gap-opening effect of the magnetic field appears to be of a purely orbital nature.
Also, contrary to the unproven experimental status of the proposal of Ref.\cite{tsvelik}, the field-induced gap predicted in Refs.\cite{dvk,kiev} may have already been 
observed in both, HOPG \cite{kopelevich} and graphene \cite{zhang}
(the over an order of 
magnitude disparity between the measured critical fields $B_0$
might stem from a comparable difference in the doping densities of the two systems).

However, the feasibility of the pairing mechanism
of Refs.\cite{dvk,kiev} was later questioned in \cite{tsvelik,son} 
whose authors argued that it would likely be completely hindered by a downward renormalization of the effective Coulomb coupling (1) due to an increase of the renormalized Fermi velocity with 
decreasing momentum \cite{gonzales}.

In what follows, we refine the analysis of Refs.\cite{dvk} and extend it to the cases of
Coulomb- and phonon-mediated Peierls- and Cooper-pairings, respectively, while accounting for the effects of the running Fermi velocity. 

{\bf Pairing instabilities and mass terms}

With an eye on a unified description of both, Peierls- and Cooper-types of pairings, we describe the Dirac fermion states as $16$-component vectors
\bea
\Psi(p)=(\psi_{C,n,\alpha}(p),\tau_2^{nm}s_2^{\alpha\beta}
\psi^\dagger_{C,m,\beta}(-p))
\nonumber\\
\psi=({1+\tau_3\over 2}+i{1-\tau_3\over 2}\otimes\sigma_2)(A_1,B_1,A_2,B_2)^{T}
\eea
where the first, second, and third indices in each of the electron- and hole-like 
$8$-component parts of this vector pertain to the sublattice ($C=A,B$), valley ($n=L,R$),
and spin ($\alpha=\uparrow,\downarrow$) degrees of freedom, respectively.

The set of operators acting in the corresponding Hilbert space is spanned by the basis of $4^4$ matrices
\be
\rho_n\otimes\sigma_a\otimes\tau_i\otimes s_\alpha
\ee
The four consecutive factors in the above tensor product act in the Nambu (particle-hole), sublattice, valley, and spin subspaces,
each of the four indices taking one of the four possible values 
$(n, a, i, \alpha=0,1,2,3)$.

With the magnetic field and chemical potential included, the Dirac fermion Green function reads
\be
\hat{G}(\varepsilon,{\bf p})=[\varepsilon-
{\hat \rho}_3\otimes(v{\bf \sigma}_\parallel{\bf p}-
\mu+{\bf s}{\bf B})+{\hat \Sigma}(\varepsilon,{\bf p})]^{-1}
\ee
where $v\approx 10^6m/s$ is the velocity of the Dirac-like nodal excitations.
The part of the self-energy $\Sigma_c=\Sigma$ commuting with the
Dirac Hamiltonian $H_0=v{\bf \sigma}_\parallel{\bf p}$ ($[H_0,\Sigma]=0$) 
accounts for the renormalization of the Dirac fermions' 
dispersion via the effective Fermi velocity $v(p)$, 
while the anticommuting one 
$\Sigma_a=\Delta$ (${\{}H_0,\Delta{\}}=0$) incorporates a (possibly, momentum-dependent) 
fermion mass function $\Delta(p)=Tr{\hat \Delta}^2/16$.

Together, they give rise to the massive Dirac fermion spectrum 
\bea
\varepsilon=\mu+\sigma B\pm E(p)\nonumber\\
E(p)={\sqrt {v^2(p)p^2+\Delta^2(p)}}
\eea
The list of putative order parameters
includes various 
Peierls- and Cooper-like fermion bilinears
\bea
\Gamma_{CDW}=\Psi^\dagger\rho_3\otimes\sigma_3\otimes\tau_{i}\otimes s_\alpha\Psi
\nonumber\\
\Gamma_{SC}=\Psi^\dagger\rho_{1,2}\otimes\sigma_0\otimes\tau_{i}\otimes s_\alpha\Psi
\eea
Among these bilinears are those which can be associated 
with $4$ different types of spin-singlet 
Dirac-type mass terms $\Psi^\dagger\Gamma_i\Psi$, classified according to
their symmetry properties under the parity ($P-$): $A(B)\leftrightarrow B(A); L(R)\leftrightarrow R(L)$ 
and time reversal ($T-$): $A(B)\leftrightarrow A(B); L(R)\leftrightarrow R(L)$ transformations:
\bea
\psi^\dagger\sigma_3\otimes\tau_1\otimes s_0\psi=A^\dagger_{L\alpha}B_{R\alpha}+
B^\dagger_{L\alpha}A_{R\alpha}+h.c.\nonumber\\
\psi^\dagger\sigma_3\otimes\tau_3\otimes s_0\psi=\sum_{i=L,R}(A^\dagger_{i\alpha}A_{i\alpha}-
B^\dagger_{i\alpha}B_{i\alpha})\nonumber\\
\psi^\dagger\sigma_3\otimes\tau_0\otimes s_0\psi=\sum_{i=L,R}sgn i
(A^\dagger_{i\alpha}A_{i\alpha}-B^\dagger_{i\alpha}B_{i\alpha})\nonumber\\
\psi^\dagger\sigma_3\otimes\tau_2\otimes s_0\psi=iA^\dagger_{L\alpha}B_{R\alpha}-
iB^\dagger_{L\alpha}A_{R\alpha}+h.c.\nonumber\\
\eea  
where we choose $sgn L(R)=\pm 1$ (hereafter, all the summations over the spin indices are implicit). 
The above operators (in the listed order) are $P,T$-even; $P$-odd, $T$-even; $P$-even, $T$-odd;
and $P,T$-odd, respectively. By replacing $s_0$ with $s_3$ one can also 
construct spin-triplet counterparts of Eqs.(7).

Notably, the above order parameters 
include both, intra- , $<A(B)^\dagger_{L,R}A(B)_{L,R}>$, and inter-node,
$<A(B)^\dagger_{L,R}B(A)_{R,L}>$, excitonic pairings  
with the total momenta $0$ and ${\vec K}_L-{\vec K}_R=2{\vec K}_L\equiv -{\vec K}_L$, respectively.
It was shown in Ref.\cite{dvk} that the former mass terms manifest themselves as a charge density wave (CDW) representing the excess/deficit of the electron density which alternates between the two sublattices.
In contrast, the latter terms correspond to the Kekule dimerization pattern which results in the tripling of the unit cell and the ensuing equivalence of the two Dirac points in the reduced Brillouin zone.

For comparison, the ordinary and staggered chemical potential/magnetic 
field terms correspond to the bilinears
\bea
\sum_{i}
\delta^{\alpha\beta}[1,{\it sgn}\alpha]
(A^\dagger_{i\alpha}A_{i\beta}+B^\dagger_{i\alpha}B_{i\beta}),\nonumber\\
\sum_i \delta^{\alpha\beta}[1,{\it sgn} \alpha](A^\dagger_{i\alpha}A_{i\beta}-B^\dagger_{i\alpha}B_{i\beta})
\eea
which only cause energy level shifts, but no gaps.

Likewise, an attractive interaction
can potentially generate the following Majorana-type mass terms
\bea
\psi\sigma_0\otimes\tau_1\otimes s_0\psi=\sum_{i=L,R}(A_{i\alpha}A_{i\beta}+B_{i\alpha}B_{i\beta})
s^{\alpha\beta}_2\nonumber\\
\psi\sigma_0\otimes\tau_2\otimes s_0\psi=\sum_{i=L,R}{\it sgn} i
(A_{i\alpha}A_{i\beta}+B_{i\alpha}B_{i\beta})s^{\alpha\beta}_2\nonumber\\
\psi\sigma_0\otimes\tau_0\otimes s_2\psi=(A_{L\alpha}B_{R\alpha}-B_{L\alpha}A_{R\alpha})\nonumber\\
\psi\sigma_0\otimes\tau_3\otimes s_0\psi=(A_{L\alpha}B_{R\alpha}-B_{L\alpha}A_{R\alpha})s_2^{\alpha\beta}
\eea  
amongst which there are both, inter-node (uniform), $<A_{L,R}B_{R,L}>$, and intra-node,
$<A(B)_{L,R}A(B)_{L,R}>$, Cooper pairings, 
the latter corresponding to a non-uniform (LOFF-type) state.

Which of the above orderings (or a combination thereof) would have a higher chance to occur, depends
on the parameter values of the overall effective electron interaction which is comprised of the direct Coulomb repulsion and the effective phonon-mediated attraction.

{\bf Electron interactions and excitonic pairing}

At small transferred momenta the effective interaction in undoped
($\mu=0$) graphene immersed in a medium with $\epsilon\sim 1$,
is dominated by the Coulomb repulsion which brings about a putative Peierls-like(excitonic) pairing instability. 
Contrary to the model problem of $N\gg 1$ Dirac fermion species studied in Ref.\cite{tsvelik} by means of the $1/N$-expansion, that of graphene features no small parameter. 

By analogy with the earlier analyses \cite{appelquist} of a fully relativistically invariant counterpart, dubbed chiral symmetry breaking, of the excitonic instability the latter can studied in the framework of the Schwinger-Dyson equation
\be
{\hat \Sigma}(\varepsilon, {\bf p})=\sum_{{\bf q}}\int {d\omega\over 2\pi}
\Gamma V(\varepsilon-\omega,{\bf p}-{\bf q})
{\omega+v{\bf \sigma}_\parallel
{\bf q}+{\hat \Sigma}(\omega,{\bf q})\over \omega^2-E^2({\bf q})+i0}
\ee
written in terms of the dressed fermion Green, interaction, and vertex ($\Gamma$) functions. 

In the case of an ultra-violet divergent 
momentum integral the very derivation of Eq.(10) becomes problematic 
because of the lack of a proper account of non-universal local terms in the Hamiltonian. While other works \cite{hubbard} resorted to 
a phenomenological approach where such terms would be treated as free parameters, we focus on the situation 
where the gap equation receives a dominant contribution
from the momenta $\Delta/v\lesssim q\ll\Lambda$, thereby justifying the derivation of Eq.(10).
Incidentally, such a behavior does occur in those situations where the fermion DOS vanishes and/or the interaction function remains sufficiently singular at small momenta. 

With the fermion polarization 
\be
\Pi(\omega, q)={Nq^2\over 16{\sqrt {v^2(q)q^2-\omega^2}}}
\ee
included, the effective interaction in Eq.(10) reads
\be
V_C(\omega,q)={2\pi gv\over q+2\pi gv\Pi(\omega, q)}
\ee
While the vertex function in Eq.(10) remains largely unaffected by the interactions, $\Gamma\approx 1$ \cite{gonzales}, the dressed fermion Green function has to incorporate, in addition to the mass term, the aforementioned renormalization of the Fermi velocity. The latter (and, concomitantly, the effective momentum-dependent coupling $g(p)$) were approximated in Ref.\cite{tsvelik,son} by a power-law 
\be
{v(p)\over v}={g\over g(p)}=({\Lambda\over p})^\eta
\ee
within the range of momenta $\Delta/v\lesssim p< \Lambda$, where $\Lambda$ is a momentum cut-off of order the inverse lattice spacing. 

Computed in the $1/N$-expansion, the exponent assumes the value \cite{son,tsvelik}
\be
\eta={8\over \pi^2N}
\ee
which remains relatively small down to the physically relevant number $N=4$ of two-component (pseudo)spinors with different values of the physical spin and valley indices.
At still lower energies and/or $g_0\ll 1$ the renormalized 
Fermi velocity shows an even slower (only logarithmic) momentum dependence \cite{tsvelik}. 

Upon neglecting the unimportant vertex corrections, switching to imaginary frequences, performing the frequency integration, and putting $\varepsilon=vp$ in Eq.(10), one arrives at the equation for the momentum-dependent mass function defined as the anticommuting part of the self-energy $\Delta(p)=\Sigma_a(ivp,{\bf p})$ 
evaluated on the mass-shell
\be
\Delta({p})=\sum_{{\bf q}}{V_C(iv(p-q),{\bf p}-{\bf q})}
{\Delta({q})\over 2E({q})}\tanh{E({q})\over 2T}
\ee
Notably, a strong momentum dependence of the integral kernel in Eq.(15) rules out the customary BCS-type solution $\Delta(p)=const$.
Concomitantly, because of the vanishing DOS of the Dirac liquid the Coulomb coupling would have to be strong enough to provide for a possible onset of pairing.

At $T=0$ and with the velocity renormalization (13) taken into account, one can cast Eq.(15) in the following approximate form
\be
\Delta(p)=\oint{d\phi\over 4\pi}
\int^\Lambda_0{q^{\eta}dq\over |{\bf p}-{\bf q}|}
{g\over (\Lambda^\eta+(\pi gN/8{\sqrt 2})|{\bf p}-{\bf q}|^\eta)}{\Delta(q)}
\ee
where $\phi$ is the angle between the vectors $\bf p$ and $\bf q$, and the additional factor of $\sqrt 2$ is characteristic of the typical energy- and momentum transfers ($\omega\approx vq$) contributing to the integral.

It is worth noting that, contrary to the naive expectation of its uniformly weakening effect on the effective momentum-dependent coupling constant, the velocity renormalization (13) affects the gap equation in a more subtle way. Namely, the above expectation is only valid
for weak couplings, whereas in the strong coupling regime $g\gg 1$
one finds a nearly complete cancellation between the velocity-dependent factors in the numerator and denominator of Eq.(16).

In what follows, we analyze the gap equation along the lines of the earlier studies of chiral symmetry breaking \cite{appelquist}. In doing so, we primarily focus on its starkly novel features that are robust enough to sustain the approximations involved. 
Namely, by following the same procedure as in 
Refs.\cite{dvk,kiev,appelquist} we, first, linearize Eq.(16) with respect to the unknown mass function and then transform it into the differential form
\be
{d^2\Delta(p)\over dp^2}+{2+\eta_N\over p}{d\Delta(p)\over dp}+{{\tilde g}(1+\eta_N)\over 2p^{2-\delta\eta}}
{\Delta(p)\over \Lambda^{\delta\eta}}=0
\ee
where $\delta\eta=\eta-\eta_N$ and the effective coupling
\be
{\tilde g}={{g}\over 1+\pi N{g}/8{\sqrt 2}}
\ee
interpolates between the weak (${\tilde g}\approx g$) and strong (${\tilde g}\approx 8{\sqrt 2}/\pi N$) coupling limits, which are attainable by putting $\eta_N=0, g\to 0$ and $\eta_N=\eta, g\to\infty$, respectively.

In order for Eq.(17) to be consistent with the original integral Eq.(15), it has be complemented with the boundary conditions
$$
 {d\Delta(p)\over dp}{|}_{p=\Delta/v}=0
$$
\be
[(1+\eta_N)\Delta(p)+p{d\Delta(p)\over dp}]{|}_{p=\Lambda}=0
\ee
The linearized Eq.(17) can also be interpreted as the radial Schroedinger equation for the zero-energy state in the
two-dimensional spherically-symmetrical potential 
$U(r)\propto 1/r^{2-\delta\eta}$. This analogy suggests that for $0\leq\delta\eta<1$ the zero-energy solution might exist for all couplings $\tilde g$ in excess of a finite threshold value ${\tilde g}_c$.

In order to illustrate the behavior of the physically relevant solution of Eq.(17) we employ the WKB method and consider a linear combination of two linearly independent functions
\be
\Delta_{\pm}(p)={C_{\pm}\over p^{1-\delta\eta/2}P(p)^{1/2}}
\exp(\pm i\int^p_\kappa P(p^\prime)dp^\prime)
\ee
where, for $p\gtrsim\kappa=[(1+\eta_N)/2{\tilde g}]^{1/\delta\eta}\Lambda$, 
\be
P^2(p)={1\over p^2}[{\tilde g}{1+\eta_N\over 2}({p\over\Lambda})^{\delta\eta}-{(1+\eta_N)^2\over 4}]
\ee
On the general grounds, one expects the physically acceptable solution to monotonically increase with decreasing momentum, eventually leveling off at a finite value $\Delta=\Delta(p)|_{p=0}$. Since the linearized equation (17) ceases to be applicable at momenta below $p\lesssim \Delta/v$, a reliable use of Eqs.(20,21) would be limited to the range of parameters $p\gtrsim\Delta/v>\kappa$. 

Strictly speaking, the applicability of the WKB technique requires the (quantized in multiples of $\pi$) values of the momentum integral in the exponential of Eq.(20) to be large, in which case the mass function $\Delta(p)$ 
develops unphysical oscillations as a function of momentum. 
Therefore, in order to find a monotonic solution which satisfies the boundary conditions (19), we choose $C_+/C_-=i$ and equate the aforementioned integral to its lowest possible 'quantized' value
\be
\int^\Lambda_\Delta{dp\over p\Lambda^{\eta/2}}
{\sqrt {{\tilde g}{1+\eta_N\over 2}(p^{\delta\eta}-\kappa^{\delta\eta})}}
+\delta_\Delta+\delta_\Lambda=\pi
\ee
where
\bea
\delta_\Lambda=tan^{-1}
{{\sqrt {2{\tilde g}(1+\eta_N)(\Lambda^{\delta\eta}-\kappa^{\delta\eta})}}\over
1+\eta_N-\delta\eta/2(1-({\tilde g}_c/{\tilde g}))}
\nonumber\\
\delta_\Delta=tan^{-1}
{{\sqrt {2{\tilde g}(1+\eta_N)(\Delta^{\delta\eta}-\kappa^{\delta\eta})}}\over
1+\eta_N+\delta\eta/2(1-(\kappa/\Delta)^{\delta\eta})}
\eea
At $\eta=0$ Eq.(17) becomes scale-invariant 
and Eq.(20) amounts to the solution obtained in Refs.\cite{dvk,kiev}
\be
\Delta(p)={\Delta^{3/2}\over (vp)^{1/2}}\sin({1\over 2}{\sqrt {2{\tilde g}-1}}\ln{\Lambda\over p}+2\delta)
\ee
where $\delta=\delta_\Lambda=\delta_\Delta=\tan^{-1}{\sqrt {2{\tilde g}-1}}$. 
It is worth noting that, regardless of the formal restrictions on the applicability of the WKB approximation, Eq.(24) remains accurate in the entire range $\Delta\lesssim p<\Lambda$ \cite{dvk,kiev}.

The maximum value of the mass 
function $\Delta=\Delta(p=\Delta/v)$ is given by Eq.(22)
\be
\Delta=v\Lambda\exp(-{2\pi-4\delta\over {\sqrt {2{\tilde g}-1}}})
\ee
Obviously, such a solution requires the coupling to be in excess of the critical value
\be
{\tilde g}_c={1\over 2}
\ee
For $N=4$, Eqs.(18) and (26) yield the critical coupling $g_c=1.13$
which appears to be tantalizingly close to (albeit still higher than) its actual value in graphene on the $SiO_2$ substrate, but below that in free-standing (suspended) graphene. Conversely, in the asymptotic $g\to\infty$ regime 
the critical number of fermion species below which the mass gets generated is $N_c=16{\sqrt 2}/\pi\approx 7.18$. 

For $\eta\ll 1$ the analysis of Eq.(22) reveals an increase in the critical value
\be
{\tilde g}_c={1\over 2}(1+\eta_N+({3\pi\delta\eta})^{2/3}+\dots),
\ee
showing that a nonzero mass can emerge even in the case of a momentum-dependent coupling.

The gap equation (15) can also be used to estimate the mean-field critical temperature. For this approximate calculation we neglect the weak (at large $N$) renormalization (13)
by putting $\eta=0$. The critical temperature can then be evaluated by solving the equation
\be
\int^\Lambda_{p_0}{dp\over p}{\sqrt {2({\tilde g}\tanh{vp\over 2T_c}-{\tilde g}_c)}}
\approx 2\pi
\ee
where $p_0$ is defined as $\tanh(vp_0/2T_c)=1/{\tilde g}$. 
As a result, Eq.(28) yields the relation between the gap and the (mean-field) critical temperature 
\be
T_c\approx{\Delta\over |\ln(1-{\tilde g}_c/{\tilde g})|},
\ee
which is markedly different from the universal ratio $T_c/\Delta=e^{C}/\pi$ characteristic of the standard BCS case. 

Despite the highly unconventional relation (29), both $\Delta$ and $T_c$ vanish rapidly upon approaching the threshold coupling.
Appyling the above rough estimates to undoped free-standing graphene one obtains the gap and the critical temperature of order $\Delta\sim T_c\approx 10eV e^{-8}\sim 4meV$ which is comparable to the gap $\sim 10meV$ reported in Ref.\cite{andrei} (an order of magnitude larger gap found in epitaxial graphene \cite{arpes} is likely to be induced by the $SiC$ substrate, though).

A finite chemical potential $\mu$ represents, depending on its sign,
a finite density of either particles or holes. It gives rise to a mismatch between the occupation numbers of the particles and holes which sets a lower limit of the momentum integration in Eq.(15). A straightforward analysis then shows that at $p\lesssim\mu/v$ 
the mass function $\Delta(p)$ tends to level off at a value below that at $\mu=0$, whereas for $p>\mu/v$ the solution obtained for $\mu=0$ remains essentially intact.

It also implies that the excitonic pairing should be sensitive to such particle-hole symmetry breaking factors as electron-hole separation ('puddles'). 
As follows from the above estimates, the presence of puddles with density 
$n\sim 10^{11}cm^{-2}$ could already hinder the excitonic gap $\Delta\sim 10meV$. 

{\bf Electron-phonon interactions and Cooper pairing}

Next, we discuss putative Cooper pairing instabilites in graphene. For interactions dominated by small momentum transfers, the onset of the Cooper instability 
would be contingent with the existence of a range of parameters 
where the effective electron interaction becomes attractive due to a more efficient screening 
of the Coulomb interaction. This situation can potentially occur, e.g., for a large-$\epsilon$ substrate (e.g., $HfO_2$) or in the presence of water molecules trapped between the graphene sheet and the substrate.

The Cooper pairing scenario would provide for an alternative explanation of the data of Ref.\cite{crommie} indicating the presence of a $\sim 100 meV$ gap at the Fermi energy in epitaxially grown graphene on a $SiC$ substrate. 
The previously proposed explanations of this finding include such mechanisms as
electronic states in the underlying buffer layer,
charging/band bending effects, coupling to the surface excitations, as well as a trivial finite size effect. As yet another idea, a phonon-mediated inelastic tunneling was argued to be behind the apparent TDOS suppression which was then misinterpreted as a gap opening \cite{wehling}.

However, considering that the system appears to be naturally doped, 
$\mu\approx 0.4 eV$ \cite{arpes},
it is conceivable that at such a sizable doping 
the Coulomb repulsion would be rather strongly screened out and the 
electron-phonon coupling could then get a chance to become dominant,
resulting in the Cooper instability. 

The phonon spectrum of graphene consists of a total of $6$ modes, which list includes (in- and out- of phase) in-plane longitudinal/transverse and out-of-plane ones. Dispersion relations of the pertinent low-energy acoustic modes with momenta near the $\Gamma$ point can be obtained from the (somewhat simplified for the sake of our qualitative discussion) elastic energy of the graphene sheet 
\bea
F={\rho\over 2}[(\partial_t{\vec u})^2+
(\partial_th)^2-\kappa^2(\partial_i^2h)^2\nonumber\\
-c^2
(\partial_iu_j+\partial_ju_i+\partial_ih\partial_jh)^2]
\eea
which governs the in-plane
phonons with a linear dispersion ($\Omega_1(q)=cq$ 
where $c\sim 10^4 m/s$) and linear coupling to the Dirac fermions
as well as the out-of-plane ('fluxor') modes
with a quadratic dispersion ($\Omega_2(q)=\kappa q^2$) and quadratic coupling \cite{oppen}. The optical modes near the $\Gamma$-point, as well as all the modes at the $K_{L,R}$-ones are gapped and their dispersions remain nearly constant ($\Omega_0(q)=\Omega_0$) at small deviations from the corresponding points in the momentum space.

The corresponding matrix elements controling the electron-phonon coupling vary with momentum as 
\be
M^a_q\propto({q\over {\sqrt m_C\Omega_a(q)}})^a
\ee
where $a=0,1,2$ and $m_C=2\times 10^{-23}g$ is the mass of carbon atom.

The overall effective electron interaction mediated by the phonon modes is given by the expression
\bea
V_{ph}(\omega,{\bf q})=\sum_{a=0,1}D_a(\omega, {\bf q})|M^a_q|^2+\nonumber\\
\sum_{{\bf q}^\prime}\int {d\omega^\prime\over 2\pi}
|M^2_{{q}^\prime}|^2
D_{2}(\omega^\prime_+, {\bf q}^\prime_+)
D_{2}(\omega^\prime_-, {\bf q}^\prime_-)
\eea
where $\omega^\prime_\pm=\omega^\prime\pm\omega/2$, 
 ${\bf q}^\prime_\pm={\bf q}^\prime\pm {\bf q}/2$, and 
\be
D_a(\omega, {\bf q})={\Omega_a(q)\over \omega^2-\Omega^2_a(q)+i0}
\ee
is the propagator of the $a$-type mode.
 
A relative importance of various phonon-induced electron interactions
can be readily assessed in the quasi-static ($\omega=0$) approximation. When treated this way, all the gapped and linear in-plane modes would have given risen to an effective interaction with the Fourier transform  
\be
V^{(0,1)}_{ph}(q)=-
\sum_{a=0,1}{|M^a_{q}|^2\over \Omega_a(q)}\sim const
\ee
which is short-ranged in the real space 
($V^{(0,1)}_{ph}(r)\propto -\delta(r)$).
By contrast, the quadratic fluxor mode would produce a 
long-ranged potential with the Fourier transform
\be
V^{(2)}_{ph}(q)=-\sum_k {|M^2_k|^2\over \Omega_2(k+q)+\Omega_2(k)}
\sim -\ln{\Lambda\over q}
\ee
which decays with the distance as $V^{(2)}_{ph}(r)\propto -1/r^2$.

Thus, depending on the relative strength of this coupling,
the fluxor mode could provide for the strongest tendency
towards the Cooper pairing, as compared to the 
gapped and linear acoustic modes.

Focusing solely on the electron-phonon coupling, one can derive the Cooper analog of the gap equation (15) for the 'on-shell' self-energy
$\Delta_{SC}(p)=\Sigma_a(vp\approx\mu,{\bf p})$.
Upon integrating over the frequencies one obtains
\be
\Delta_{SC}(p)=\sum_{a, {\bf q}}
{|M^a_{{\bf p}-{\bf q}}|^2
\over {\Omega_a({{\bf p}-{\bf q}})+E(q)}}
{\Delta_{SC}(q)\over 2E(q)}\tanh{E(q)\over 2T}
\ee
Contrary to the case of the excitonic instability at $\mu=0$, the solution of Eq.(36) exists for an arbitrarily weak electron-phonon interaction (which, however, is assumed to be stronger than the screened Coulomb one). This solution appears to be essentially independent of the momentum for $p\lesssim \mu/v$, and the gap conforms, by and large, 
to the standard BCS formula
\be
\Delta_{SC}\approx\sum_{a=0,1,2}E_ae^{-1/\lambda_{a}}
\ee
where the characteristic energy scales and dimensionless coupling constants corresponding to the 
different modes are given by the expressions
\bea
E_0\sim min[\Omega_0,\mu],~~~~E_1\sim {c\mu\over v},~~~~E_2\sim \mu({\Lambda\kappa\over v})^{1/2}
\eea
and
\bea
\lambda_0\sim \mu {D^2a^2\over {m\Omega^2_0v^2}},~~~~
\lambda_1\sim {\mu\over mc},\nonumber\\
\lambda_2\sim {\mu a^2 \over m^2\kappa^3}\ln
{v^3\Lambda\over \kappa\mu^2},
\eea
respectively. Here $D\approx 7 eV/A$ is the deformation potential, $a=1.42A$ is the lattice spacing, and $\Omega_0\sim 0.2 eV$ is the energy of the Raman-active gapped phonon modes.

Contrary to excitonic pairing, the Cooper one is facilitated by an increasing chemical potential 
(of either sign), 
while it gets suppressed by a Zeeman-coupled magnetic field.
Also, the logarithmic factor in the coupling to the fluxor mode $\lambda_2$ is another manusfestation of the long-ranged nature of the effective electron interaction (35) mediated by this mode.  

As far as the actual values of the couplings (39) are concerned, those predicted by the density functional theory \cite{louie} seem to be much 
too weak to explain the experimentally observed downward renormalization of the Fermi velocity \cite{arpes,andrei,crommie}.
Specifically, apart from being of the sign opposite to that of the predicted Coulomb-only renormalization (13), the experimentally measured velocity change appears to be quite large in magnitude ($\approx 20\%$) and suggestive of the 
effective electron-phonon coupling of order 
$\lambda_0\sim 0.2-0.4$.

The analysis of Ref.\cite{basko} limited to 
the gapped in-plane modes
indicates that, depending on the mode's symmetry, the corresponding coupling constant might become strongly enhanced at low momenta by the Coulomb interactions. Namely, in Ref.\cite{basko} the coupling constant for the
optical $'A1'$-mode at the $K$-point 
was found to increase by over an order of magnitude
from its microscopic value as a result of the scaling 
$\lambda_0(p)=\lambda_0({v(p)/v})^2$. 

For the largest estimated values of the above parameters
the gap can be as high as $\Delta\approx 0.4eV e^{-2.5}\sim 30meV$ which compares favorably with the experimentally observed $\sim 100meV$ gap \cite{crommie}.

The lack of a significant momentum dependence in Eq.(36) would give rise to the BCS relation between $\Delta_{SC}$ and the mean-field critical temperature $T^{SC}_c$. 
Although the above estimates suggest that for the 
experimentally relevant electron densities $n\sim 10^{13}cm^{-2}$ the latter could be as high as $T^{SC}_c\sim 150K$, the prospects of observing the Cooper instability in 
graphene are likely to be hampered by a number of factors.

Firstly, the above Peierls- and Cooper-pairings would primarily manifest themselves through a "pseudogap" in the electron spectrum which emerges below the corresponding mean-field critical temperatures. However, a true transition in such a two-dimensional system as graphene would only take place at a temperature determined by the (un)binding of the vortex-antivortex excitations of the order parameter in question.

The critical temperature of the underlying Kosterlitz-Thouless (KT) transition is related to the rigidity with respect to the fluctuations of that order parameter. The latter can be computed by expanding the pertinent susceptibility 
\bea
\rho_s={\Pi_{ij}(0,q)\over q_iq_j}|_{q\to 0}=\nonumber\\
\Delta^2\sum_p\int {d\omega\over 2\pi}
Tr{\hat \Gamma}_i{\hat G}(\omega, p+q){\hat \Gamma}_j
{\hat G}(\omega, p)\approx {1\over 4\pi}\Delta
\eea
to second order in the center-of-mass momentum $q$, which procedure yields the KT temperature
\be 
T_{KT}={\pi\over 2}\rho_s\approx {1\over 8}\Delta
\ee
The observation that the KT transition temperature is smaller than the mean-field one by an order of magnitude is in agreement with the previous work of Refs.\cite{macdonald}.

Albeit being robust against potential disorder, the onset of the Cooper pairing can be hindered by a rippling of the graphene sheet. The latter was argued to produce effective 
magnetic fields as high as $B\sim 1-5 T$ \cite{ripples}, whose pair-breaking effect could potentially eliminate the propensity towards the Cooper instability altogether,
since for $\Delta_{SC}\sim 0.1eV$ the upper 
critical field $B_{c2}=\Phi_0\Delta^2_{SC}/2\pi v^2$ would be of order $\sim 4T$.

{\bf Conclusions}

In summary, we carried out the analysis of excitonic (Peierls) and Cooper pairing instabilities in graphene which presents a physically relevant example of the nodal fermion systems with a power-law DOS and  long-range pairwise interactions. 

We find that undoped graphene can be prone to the onset of a variety of excitonic instabilities, provided that the Coulomb coupling is sufficiently (albeit not unphysically) strong.
Unlike in the low-energy (hence, weak-coupling) BCS-like pairing scenaria of Refs.\cite{tsvelik,macdonald}, the non-BCS solution of the gap equation Eq.(15)
is strongly momentum-dependent and requires the coupling 
to be in excess of a finite threshold value $g_c$.
Moreover, the perturbative downward renormalization of the 
running coupling constant $g(p)$ does not appear to have a dramatic impact on the onset of this excitonic instability (or a lack thereof). 

With increasing doping, the excitonic instability 
becomes suppressed, while a propensity towards the zero-threshold Cooper pairing driven by the phonon-mediated attraction gets stronger
(in practice, however, it remains countered by the residual screened Coulomb repulsion).
Should either of these instabilities fully develop, it would give rise to a momentum-dependent gap function and/or other experimentally testable non-BCS features.

A better understanding of the phenomena resulting in a finite gap in the nodal fermion spectrum is going to be instrumental for the proposed applications of graphene in post-silicon electronics. Therefore, one should expect that future experiments will help to ascertain the actual status of the above scenarios before long. 
 
The author acknowledges the hospitality at the Aspen Center for Physics where this work was completed. This research was supported by NSF under Grant DMR-0349881.

{\it Note added}: \\
After this work was first made public (arXiv:0807.0676), the results of two new Monte Carlo simulations were 
released, according to which the aforementioned excitonic transition in free-standing 
graphene might occur at the critical coupling as low as $g_c=1.1$ for $N=4$
\cite{drut} and the number of fermions
as high as $N_c=9.4$ for $g\to\infty$ \cite{hands} (note that in both preprints the quoted number of fermions $N=2$ pertains to the {\it four}-component bi-spinors, which corresponds to $N=4$ in this paper).
Also, in the recent work \cite{liu} the integral gap equation (10) was solved numerically without any further approximations 
and the critical number of fermions was found to be $N_c\approx 7$.
All the above values appear to be remarkably close to those obtained in this paper.

%\end{thebibliography}
%\end{multicols}

\end{document}